\documentclass[pre,twocolumn,aps,showpacs]{revtex4}
\newcommand{\bec}[1]{\mbox{\boldmath $ #1$}}
\begin{document}
\centerline{\bf PHYSICAL REVIEW E, v. 65, 036303 (2002)}
\bigskip
\bigskip
\title{Magnetic fluctuations with a zero mean field in
a random fluid flow with a finite correlation time and a small
magnetic diffusion}
\author{Nathan Kleeorin}
\email{nat@menix.bgu.ac.il}
\author{Igor Rogachevskii}
\email{gary@menix.bgu.ac.il}
\homepage{http://www.bgu.ac.il/~gary}
\affiliation{Department of Mechanical Engineering, The Ben-Gurion
University of the Negev, \\
POB 653, Beer-Sheva 84105, Israel}
\author{Dmitry Sokoloff}
\email{sokoloff@dds.srcc.msu.su}
\homepage{http://www.srcc.msu.su/lemg} \affiliation{ Department of
Physics,  Moscow State University, Moscow 117234, Russia}
\date{Received 6 August 2001; published 12 February 2002}
\begin{abstract}
Magnetic fluctuations with a zero mean field in a random flow with
a finite correlation time and a small yet finite magnetic
diffusion are studied. Equation for the second-order correlation
function of a magnetic field is derived. This equation comprises
spatial derivatives of high orders due to a non-local nature of
magnetic field transport in a random velocity field with a finite
correlation time. For a random Gaussian velocity field with a
small correlation time the equation for the second-order
correlation function of the magnetic field is a third-order
partial differential equation. For this velocity field and a small
magnetic diffusion with large magnetic Prandtl numbers the growth
rate of the second moment of magnetic field is estimated. The
finite correlation time of a turbulent velocity field causes an
increase of the growth rate of magnetic fluctuations. It is
demonstrated that the results obtained for the cases of a small
yet finite magnetic diffusion and a zero magnetic diffusion are
different. Astrophysical applications of the obtained results are
discussed.
\end{abstract}

\maketitle
\section{Introduction}

In recent time magnetic fluctuations are a subject of intensive study
(see, e.g.,
\cite{ZMR88,ZRS90,CG95,KA92,GD94,KRA94,GS96,RK97,KR99,SK01}).
There are two types of magnetic fluctuations: the fluctuations
with a zero and a nonzero mean magnetic field.
These two types of magnetic fluctuations have
different mechanisms of generation and different properties. Magnetic
fluctuations with a zero mean magnetic field in a random velocity field
are generated by the stretch-twist-fold mechanism (see, e.g.,
\cite{ZMR88,ZRS90}). On the other hand,
magnetic fluctuations with a nonzero mean magnetic field are
generated by a tangling of the mean magnetic field by a random
velocity field (see, e.g., \cite{M78,P79,KR80,ZRS83}).

In the present paper we considered only magnetic fluctuations with a zero mean
magnetic field which where observed, {\em e.g.,} in the ionosphere of Venus
(see, e.g., \cite{RE79,KRE94}), in the quiet sun (see, e.g., \cite{ZRS83})
and probably in galaxies (see, e.g., \cite{RSS88}).
In spite of that the dynamics of a mean magnetic
field at least in kinematic (linear) stage is well studied
(see, e.g., \cite{M78,P79,KR80,ZRS83,RSS88}), a generation of
magnetic fluctuations with a zero mean magnetic field even in kinematic stage
still remains a subject of numerous discussions. Most studies starting
with a seminal paper by Kazantsev \cite{K68} were performed in the
delta-correlated in time approximation for a random velocity field
(see, e.g., \cite{ZMR88,ZRS90,RK97,KR99}, and references therein).
A use the delta-correlated in time approximation for a random velocity field
is a great mathematical convenience.

However, a real velocity field in astrophysical and geophysical
applications cannot be considered as the delta-correlated in time
velocity field. As follows from the analysis in \cite{DM84,EKRS99}
a finite correlation time of the velocity field does not essentially change a
form of the mean-field equations and the growth rates of the mean fields.
In particular, there is a wide range of scales in which the mean-field
equations are the second-order partial differential equations (in spatial
derivatives). However, the effect of a finite correlation time of the
velocity field on magnetic fluctuations is poorly understood. It is not clear
how conditions for the generation of magnetic fluctuations are changed
in a random velocity field with a finite correlation time.

In this study we took into account a finite correlation time
of a random velocity field and a small yet finite
magnetic diffusion caused by an electrical conductivity of fluid.
We derived an equation for the second-order correlation function
of magnetic field in a random velocity field with a finite
correlation time using a method described in \cite{DM84,EKRS99,EKRS00}.
The derived equation comprises spatial derivatives of high
orders. For a random Gaussian velocity field with a small correlation time
the equation for the second-order correlation function of the magnetic field
is a third-order partial differential equation.
We calculated the growth rate of the second moment of magnetic field
for this velocity field and a small magnetic diffusion with large magnetic
Prandl numbers. In the limit of extremely small correlation time of a random
velocity field we recovered the results obtained in the delta-correlated
in time approximation for a random velocity field.

Recently, the finite correlation time effects of a random velocity field
in the kinematic dynamo in the case of a zero magnetic diffusion have been
studied in \cite{SK01}. We will show that the results obtained for the cases
of a zero magnetic diffusion and of a small yet finite magnetic diffusion
are different.

\section{Governing equations}

We study magnetic fluctuations with a zero mean
magnetic field. A mechanism of the generation of
magnetic fluctuations with a zero mean magnetic field was proposed by
Zeldovich (see, e.g.,  \cite{ZMR88,ZRS90}) and comprises stretching, twisting
and folding of the original loop of a magnetic field. These non-trivial
motions are three dimensional and result in an amplification of the
magnetic field. The magnetic field $ {\bf b}(t,{\bf r}) $ is determined
by the induction equation
\begin{eqnarray}
{\partial {\bf b} \over \partial t}  + ({\bf v} \cdot \bec{\bf \nabla})
{\bf b} = ({\bf b} \cdot \bec{\bf \nabla}) {\bf v} - {\bf b}
(\bec{\bf  \nabla} \cdot {\bf v}) + D_{m} \Delta {\bf b} \;,
\label{T1}
\end{eqnarray}
where $ D_{m} $ is the magnetic diffusion caused by an electrical
conductivity of a fluid, $  {\bf v} $ is a random velocity
field. The goal of the present paper is to derive equation for the
second-order correlation function of the magnetic field in a random
velocity field with a finite correlation time.

Now we discuss a method of derivation of the equation for the
second-order correlation function of the magnetic field
(for details, see Appendix A).
We use an exact solution of Eq. (\ref{T1}) in the form of a
functional integral for an arbitrary velocity field taking into account
a small yet finite molecular magnetic diffusion. The molecular magnetic
diffusion can be described by a random Brownian motions of a particle.
The functional integral implies an averaging over a random Brownian
motions of a particle. The form of the exact solution used in the present
paper allows us to separate the averaging over both, a random Brownian
motions of a particle and a random velocity field.
This method allows us to derive equation for the second-order correlation
function $ \Phi_{ij}(t, {\bf x}, {\bf y}) =  \langle b_{i}(t,{\bf x})
b_{j}(t, {\bf y}) \rangle $ of the magnetic field:
\begin{eqnarray}
&& \Phi_{ij}(t, {\bf r}) = P_{ijpl}(\tau,{\bf r}, i \bec{\nabla})
\Phi_{pl}(s, {\bf r}) \;,
\label{A51}\\
&& P_{ijpl}(\tau,{\bf r}, i \bec{\nabla}) = M_{\bec{\xi}} \{
\langle G_{ip}({\bf x}) G_{jl}({\bf y}) \exp(\bec{\tilde \xi}
\cdot \bec{\nabla}) \rangle \} \label{A52}
\end{eqnarray}
(see Appendix A), where $ \tau = t - s ,$ $ \, G_{ij}({\bf x})
\equiv G_{ij}(t,s, \bec{\xi}({\bf x})) $ is determined by equation
$ d G_{ij}(t,s,\bec{\xi}) / ds = N_{ik} G_{kj}(t,s,\bec{\xi}) $
with  the initial condition $ G_{ij}(t=s) = \delta_{ij} ,$ and the
tensor $ G_{ij} $ can be considered as the Jacobian for magnetic
field transport. Here $ N_{ij} = \partial v_{i} /
\partial x_{j} - \delta_{ij} (\bec{\bf  \nabla} \cdot {\bf v}) ,$
$ \, M_{\bec{\xi}} \{ \cdot \} $ denotes the mathematical expectation
over the Wiener paths $ \bec{\xi}({\bf x}) = {\bf x} - \int_{0}^{t-s}
{\bf v}(t-\sigma,\bec{\xi}) \,d\sigma + (2 D_{m})^{1/2} {\bf w}(t-s) ,$
and $ \bec{\tilde \xi} = \bec{\xi}({\bf y})
- \bec{\xi}({\bf x}) - {\bf r} ,$ $ \, {\bf r} = {\bf y} - {\bf x} ,$
$ \, \bec{\nabla} = \partial / \partial {\bf r} ,$
the angular brackets $ \langle \cdot \rangle $
denote the ensemble average over the random velocity field,
and the molecular magnetic diffusion, $ D_{m} ,$ is described by a Wiener
process $ {\bf w}(t) .$ Another equivalent approach which includes a weak
molecular diffusion in a Lagrangian map, with a Green's function
was considered in \cite{V88,V89}.

  Equation (\ref{A51}) for the second moment of a magnetic field comprises
spatial derivatives of high orders due to a non-local nature of
turbulent transport of magnetic field in a random velocity field with a
finite correlation time (for details, see Appendix A).

\section{The random Gaussian velocity field with a small correlation time}

Now we use the model of the random Gaussian velocity field with a small
yet finite correlation time. We seek a solution for the second moment
of the magnetic field in the form
\begin{eqnarray}
\Phi_{ij}(t,{\bf r}) &\equiv& \langle b_i(t,{\bf x}) b_j(t,{\bf
y}) \rangle = W(t,r) \delta_{ij}
\nonumber\\
& & + (r W' / 2) P_{ij}({\bf r}) \;, \label{D7}
\end{eqnarray}
where $ W(t,r) = \langle \tilde b(t,{\bf x}) \tilde b(t,{\bf y})
\rangle ,$ $ \, \tilde b = {\bf b} \cdot {\bf r} ,$
$ \, {\bf r} = {\bf y} - {\bf x} ,$
$ \, P_{ij}({\bf r}) = \delta_{ij} - r_{ij} ,$
$ \, r_{ij} = r_{i} r_{j} / r^{2} $ and
$ W' = \partial W(t,r) / \partial r .$ This form of the second
moment corresponds to the condition $ \bec{\nabla} \cdot {\bf b} = 0 $
and an assumption of the homogeneous and isotropic magnetic fluctuations.
We considered a homogeneous, isotropic and incompressible random velocity
field (see below). The equation for the correlation function
$ W(t,r) $ is given by
\begin{eqnarray}
{\partial W(t,r) \over \partial t} &=& (1/3) \, \sigma_{_{\xi}} \,
r^{3} \, W''' + m^{-1}(r) \, W''
\nonumber\\
& & + \mu(r) \, W' + \kappa \, W \;, \label{D8}
\end{eqnarray}
(for details, see Appendix B), where in the leading order of
asymptotic expansion
$ \kappa = (20/3) (1 + \sigma_{_{\xi}}/4) ,$ and
\begin{eqnarray*}
1 / m(r) &=& 2 / \Pr + (2/3) r^{2}(1 + 8 \sigma_{_{\xi}}) \;,
\\
\mu(r) &=&  {4 \over m(r) r} + \biggl( {1 \over m(r)} \biggr)' - 27
\, \sigma_{_{\xi}} \, r  \;,
\end{eqnarray*}
$ \Pr = \nu / D_{m} $ is the magnetic Prandtl number,
$ \nu $ is the kinematic viscosity, $ \sigma_{_{\xi}} = (2/3) {\rm St}^{2} ,$
$ \, {\rm St} = \tau u_{d} / l_{d} $ is the Strouhal number.
Equation (\ref{D8}) is written in dimensionless form: the distance $ r $ is
measured in the units of the inner scale of turbulence
$ l_{d} = l_{0} {\rm Re}^{-3/4} ,$ the time $ t $ is measured in the units
$ \tau_{d} = \tau_{0} {\rm Re}^{-1/2} ,$ where $ \tau_{d} $ is the turnover
time of eddies in the inner scale $ l_{d} $ and the velocity $ v $ is
measured in the units $ u_{d} = l_{d} / \tau_{d} ,$
$ {\rm Re} = u_0 l_0 / \nu \gg 1 $ is the Reynolds number,
$ u_{0} $ is the characteristic turbulent velocity in the maximum scale
of turbulent motions $ l_{0} $ and $ \tau_{0} = l_{0} / u_{0} .$
In this study we consider the case of large magnetic Prandtl numbers.
For the derivation of Eq. (\ref{D8}) we used a
homogeneous, isotropic and incompressible random velocity field
and the correlation function $ f_{ij}(t,{\bf r}) = \langle v_{i}(t,{\bf x})
v_{j}(t,{\bf y}) \rangle $ for the velocity field is given by
\begin{eqnarray}
f_{ij} = (1/3) [F(r) \delta_{ij} + (r F' / 2) P_{ij}({\bf r})] \; .
\label{D6}
\end{eqnarray}
We assumed that in dissipative range $ (0 \leq r \leq 1) $ of a turbulent
velocity field the function $ F(r) $ is given by $ F(r) = 1 - r^{2} .$

Now we analyze a solution of Eq. (\ref{D8}). In a molecular magnetic diffusion region of
scales whereby $ r \ll \Pr^{-1/2}$, all terms $\propto r^2$
may be neglected. Then the solution of Eq. (\ref{D8}) is given by
$ W(t,r) = (1 - \alpha \, \Pr \, r^{2}) \exp(\gamma t)$, where
$ \gamma $ are the eigenvalues to be found, $ \alpha
= (\kappa - \gamma) / 20 $ and $\kappa > \gamma .$
In a turbulent magnetic diffusion region of scales, $\Pr^{-1/2} \ll
r \ll 1$, the molecular magnetic diffusion term $\propto 1/\Pr$
is negligible. Thus, the solution of Eq. (\ref{D8}) in this region is
$ W(t,r) = A_{1} r ^{-\lambda} \exp(\gamma t)$, where $ \lambda $ is
determined by an equation
\begin{eqnarray}
\sigma_{_{\xi}} \lambda^{3} &-&  (3 + 13 \sigma_{_{\xi}})
\lambda^{2} + (15 - 1226 \sigma_{_{\xi}}) \lambda
\nonumber\\
&+& (9/2) \gamma - 30 - 5 \sigma_{_{\xi}} = 0 \; . \label{D9}
\end{eqnarray}
For a small parameter $\sigma_{_{\xi}}$ we obtain $ \lambda \approx 5/2
- 424 \, \sigma_{_{\xi}} \pm i a_{0} ,$ where $ a_{0}^{2} =
3 (5 - 2 \gamma_{0}) / 4 ,$ $ \gamma =
\gamma_{0} + \sigma_{_{\xi}} \gamma_{1} $ and $ \gamma_{1} \approx 348 .$
For $r \gg 1$ the solution for $W(t,r)$ decays rapidly with $r .$
The value $ \gamma_0$ can be calculated by
matching the correlation function $ W(t,r) $ and its first and second
derivatives at the boundaries of the above regions, i.e.,
at the points $r = \Pr^{-1/2}$ and $r = 1$. In particular,
the matching yields $ a_{0} \approx 2 \pi k / \ln \Pr ,$
where the parameter $ k = 1; 2; 3; \ldots $ determines modes with different
numbers of zero-points $ (W = 0) $ in the correlation function $ W(r) .$
In particular, the mode with $ k=1 $ has only one zero-point in the
correlation function $ W(r) .$ Thus, the growth rate $ \gamma $ of magnetic
fluctuations is given by
\begin{eqnarray}
\gamma \approx {5 \over 2} - {2 \over 3} \biggl({2 \pi k \over
\ln \Pr} \biggr)^{2} + 348 \, \sigma_{_{\xi}} \; .
\label{D10}
\end{eqnarray}
The correlation function $  W(t,r) $ has global maximum at $ r = 0 .$
This implies that the real part of $ \lambda $ is positive. Thus,
$ \tau < 0.1 \, (l_{d} / u_{d}) .$
It follows from Eq. (\ref{D10}) that the finite
correlation time of a turbulent velocity field causes an increase
of the growth rate of magnetic fluctuations. The latter is important
in view of applications in astrophysics and planetary physics because
the real velocity field has a finite correlation time. Note that the
considered case corresponds to the fast dynamo because the growth
rate tends to the nonzero constant at very large magnetic Reynolds
numbers.

\section{Discussion}

In the present paper we studied an effect of a finite correlation time
of a turbulent velocity field on dynamics of magnetic fluctuations
with a zero mean magnetic field in the case of a small yet finite
magnetic diffusion. The finite correlation time
results in an increase of the growth rate of magnetic fluctuations.
However, the developed theory is limited by an assumption about small
correlation time, {\em i.e.,} $ \tau < 0.1 \, (l_{d} / u_{d}) .$
The latter estimate is quit realistic {\em e.g.,} for galactic
turbulence (see Ref. \cite{RSS88}).
We showed also that for an arbitrary  correlation time
of a turbulent velocity field the equation for the second moment of
the turbulent magnetic field comprises higher-order spatial derivatives.

In this study we took into account a small yet finite magnetic
diffusion caused by an electrical conductivity of a fluid.
The obtained results are different from that derived for a zero
magnetic diffusion (see \cite{SK01}). In particular, the
finite correlation time of a turbulent velocity field
reduces the growth rate of magnetic fluctuations in the case
of a zero magnetic diffusion (see \cite{SK01}). A difference between
two cases with a zero magnetic diffusion and a small yet finite magnetic
diffusion can be demonstrated even for the $\delta$-correlated in
time random velocity field. For instance, for large magnetic Prandtl
numbers the growth rate of the second moment of a turbulent
magnetic field is given by
\begin{eqnarray}
\gamma = {5 (1 + \sigma/3) \over 2(1 + 3 \sigma)} -
{2 (1 + 3 \sigma) \over 3(1 + \sigma)} \biggl({2 \pi k \over
\ln \Pr} \biggr)^{2} \;,
\label{D12}
\end{eqnarray}
where $ \sigma = \langle (\bec{\nabla} \cdot {\bf v})^{2} \rangle /
\langle (\bec{\nabla} \times {\bf v})^{2} \rangle $ is the degree of
compressibility of fluid velocity field.
Equation (\ref{D12}) is obtained using Eqs. (29) and (30) in Ref.
\cite{RK97} and implies that the compressibility of fluid velocity
field causes a reduction of the growth rate of the second moment of a
turbulent magnetic field. On the other hand, in the case of a zero magnetic
diffusion the growth rate of the second moment of
magnetic fluctuations generated by the $\delta$-correlated in
time random velocity field is given by
\begin{eqnarray}
\gamma = {10 (1 + 2\sigma) \over 3(1 + \sigma)} \;
\label{D14}
\end{eqnarray}
(see \cite{SK01}), and the compressibility results in an increase the
growth rate of the second moment of a turbulent magnetic field.
This implies that a transition from the case of a zero magnetic
diffusion to that of a small yet finite magnetic diffusion is
singular. The limit of zero magnetic diffusion is singular because the
growth rate $ \gamma $ of magnetic fluctuations is discontinuous at
zero magnetic diffusion, {\em i.e.,} it is different from the limit
of magnetic diffusion tending to zero. This stresses a danger for an
application of the results obtained for a zero magnetic diffusion to
astrophysics and planetary physics where the magnetic diffusion caused by an
electrical conductivity of fluid is small yet finite.

\begin{acknowledgments}
We acknowledge support from INTAS Program Foundation (Grant No. 99-348)
and RFBR grant 01-02-16158.
DS is grateful to a special fund of the Faculty of Engineering of the
Ben-Gurion University of the Negev for visiting senior scientists.
\end{acknowledgments}

\appendix
\section{Derivation of Eq. (\ref{A51})}

When $ D_{m} \not= 0 $ the magnetic field $ {\bf b}(t, {\bf x}) $ is given by
\begin{eqnarray}
b_{i}(t, {\bf x}) = M_{\bec{\xi}} \{G_{ij}(t,\bec{\xi}) \,
\exp(\bec{\xi}^{\ast} \cdot \bec{\nabla}) b_{j}(s, {\bf x}) \} \;,
\label{A5}
\end{eqnarray}
where $ \bec{\xi}^{\ast} = \bec{\xi} - {\bf x} .$ In order to derive
Eq. (\ref{A5}) we use an exact solution of Eq. (\ref{T1}) with an initial
condition $ {\bf b}(t=s,{\bf x}) = {\bf b}(s,{\bf x}) $ in the form
of the Feynman-Kac formula:
\begin{eqnarray}
b_{i}(t,{\bf x})  = M_{\bec{\xi}} \{G_{ij}(t,s,\bec{\xi}(t,s)) \,
b_{j}(s,\bec{\xi}(t,s))\} \;,
\label{T5}
\end{eqnarray}
where $ d G_{ij}(t,s,\bec{\xi}) / ds = N_{ik}
G_{kj}(t,s,\bec{\xi}) ,$ $ \, N_{ij} = \partial v_{i} / \partial
x_{j} - \delta_{ij} (\bec{\nabla} \cdot {\bf v}) $ and $ \,
M_{\bec{\xi}} \{ \cdot \} $ denotes the mathematical expectation
over the Wiener paths $ \bec{\xi}(t,s) = {\bf x} - \int_{0}^{t-s}
{\bf v}[t-\sigma,\bec{\xi}(t,\sigma)] \,d\sigma + (2 D_{m})^{1/2}
{\bf w}(t-s) .$ Now we assume that
\begin{eqnarray}
{\bf b}(t, \bec{\xi}) = \int \exp(i \bec{\xi} \cdot {\bf q})
{\bf b}(s, {\bf q}) \,d{\bf q} \; .
\label{CC8}
\end{eqnarray}
Substituting Eq. (\ref{CC8}) into Eq. (\ref{T5}) we obtain
\begin{eqnarray}
b_{i}(s, {\bf x}) &=& \int M_{\bec{\xi}}
\{G_{ij}(t,s,\bec{\xi}(t,s)) \, \exp[i \bec{\xi}^{\ast} \cdot {\bf
q}] \, b_{j}(s, {\bf q}) \}
\nonumber\\
& & \times \exp(i {\bf q} \cdot {\bf x}) \,d{\bf q} \; .
\label{C8}
\end{eqnarray}
In Eq. (\ref{C8}) we expand the function
$ \exp[i \bec{\xi}^{\ast} \cdot {\bf q}] $ in Taylor series at
$ {\bf q} = 0 ,$ i.e.,
$ \exp[i \bec{\xi}^{\ast} \cdot {\bf q}] = \sum_{k=0}^{\infty}
(1/k!) (i \bec{\xi}^{\ast} \cdot {\bf q})^{k} .$
Using the identity $ (i {\bf q})^{k} \exp[i {\bf x} \cdot {\bf q}] =
\bec{\nabla}^{k} \exp[i {\bf x} \cdot {\bf q}] $
and Eq. (\ref{C8}) we get
\begin{eqnarray}
b_{i}(t, {\bf x}) &=& M_{\bec{\xi}} \{G_{ij}(t,s,\bec{\xi})
[\sum_{k=0}^{\infty} (1/k!) (\bec{\xi}^{\ast} \cdot
\bec{\nabla})^{k}]
\nonumber\\
& & \times \int b_{j}(s, {\bf q}) \exp(i {\bf q} \cdot {\bf x})
\,d{\bf q} \} \; . \label{BC8}
\end{eqnarray}
After the inverse Fourier transformation in Eq. (\ref{BC8}) we
obtain Eq. (\ref{A5}). Equation (\ref{CC8}) can be formally considered
as an inverse Fourier transformation of the function $ b_{i}(t,
\bec{\xi}) .$ However, $ \bec{\xi} $ is the Wiener path which is not
a usual spatial variable. Therefore, it is desirable to derive
Eq. (\ref{A5})  by a more rigorous method as it is done below.

To this end we use an exact solution
of the Cauchy problem for Eq. (\ref{T1}) with an initial condition $
{\bf b}(t=s,{\bf x}) = {\bf b}(s,{\bf x}) $ in the form
\begin{eqnarray}
b_{i}(t,{\bf x}) = M_{\bec{\zeta}} \{J(t,s,\bec{\zeta}) \tilde
G_{ij}(t,s,\bec{\zeta}) \, b_{j}(s,\bec{\zeta}(t,s)) \} \;,
\label{T2}
\end{eqnarray}
where the matrix $ \tilde G_{ij} $ is determined
by the equation
$ d \tilde G_{ij}(t,s,\bec{\zeta}) / d s = N_{ik}
\tilde G_{kj}(t,s,\bec{\zeta}) $
with the initial condition $ \tilde G_{ij}(t=s) =
\delta_{ij} ,$ and the function $ J(t,s,\bec{\zeta}) $ is given by
\begin{eqnarray}
J(t,s,\bec{\zeta}) = \exp  [- (2 D_{m})^{-1/2}
\nonumber\\
\times \int_{0}^{t-s} {\bf v}(t-\eta,\bec{\zeta}(t,\eta)) \cdot
\,d{\bf w}(\eta)
\nonumber \\
- (4 D_{m})^{-1} \int_{0}^{t-s} {\bf
v}^{2}(t-\eta,\bec{\zeta}(t,\eta)) \,d{\eta} ] \;, \label{T4}
\end{eqnarray}
$ {\bf w}(t) $ is a Wiener process, and
$ M_{\bec{\zeta}} \{ \cdot \} $ denotes the mathematical expectation over
the paths $ \bec{\zeta}(t,s) = {\bf x} + (2 D_{m})^{1/2} ({\bf w}(t) -
{\bf w}(s)) .$ The solution (\ref{T2}) was first found
in \cite{DM84} for a magnetic field in an incompressible fluid flow.
Equation (\ref{T2}) generalizes the solution obtained in \cite{DM84} for a
magnetic field in a compressible random velocity field.
The first integral $ \int_{0}^{t-s}
{\bf v}(t-\eta,\bec{\zeta}(t,\eta)) \cdot \,d{\bf w}(\eta) $
in Eq. (\ref{T4}) is the Ito stochastic integral (see, e.g., \cite{Mc69}).

The difference between the solutions (\ref{T2}) and (\ref{T5})
is as follows. The function $ b_{j}(s,\bec{\xi}(t,s)) $ in Eq. (\ref{T5})
explicitly depends on the random velocity field $ {\bf v} $ via
the Wiener path $ \bec{\xi} ,$ while the function
$ b_{j}(s,\bec{\zeta}(t,s)) $ in
Eq. (\ref{T2}) is independent of the velocity $ {\bf v} .$
Trajectories in the Feynman-Kac formula (\ref{T5}) are determined by both,
a random velocity field and magnetic diffusion. On the other hand,
trajectories in Eq. (\ref{T2}) are determined only by magnetic diffusion.
Due to the Markovian property of the Wiener process the solution
(\ref{T2}) can be rewritten in the form
\begin{eqnarray}
b_{i}(t,{\bf x}) &=& E \{S_{ij}(t,s,{\bf x},{\bf X}')
\, b_{j}(s,{\bf X}') \}
\nonumber\\
& = & \int Q_{ij}(t,s,{\bf x},{\bf x}') b_{j}(s,{\bf x}') \,d {\bf
x}' \;, \label{T8}
\end{eqnarray}
where
\begin{eqnarray}
Q_{ij}(t,s,{\bf x},{\bf x}')  &=& [4 \pi D_{m} (t - s)]^{3/2}
\exp \biggl(- {({\bf x}' - {\bf x})^{2} \over 4 D_{m} (t - s) } \biggr)
\nonumber\\
& & \times S_{ij}(t,s,{\bf x},{\bf x}') \;, \label{T9}
\end{eqnarray}
$ S_{ij}(t,s,{\bf x},{\bf x}') = M_{\bec{\mu}} \{J(t,s,\bec{\mu})
\tilde G_{ij}(t,s,\bec{\mu}) \} $ and $ \, M_{\bec{\mu}} \{ \cdot \} $ means
the path integral taken over the set of trajectories $ \bec{\mu} $ which
connect points $ (t,{\bf x}) $ and $ (s,{\bf x}') .$  The mathematical
expectation $ E \{ \cdot \} $ in Eq. (\ref{T8}) denotes the averaging
over the set of random points $ {\bf X}' $ which have a Gaussian
statistics (see, e.g., \cite{S80}). We used here the following property
of the averaging over the Wiener process
$ E \{ M_{\bec{\mu}} \{ \cdot \} \} =  M_{\bec{\zeta}} \{ \cdot \} .$
We considered a random velocity field with a finite renewal time.
In the intervals $ \ldots (- \tau, 0]; (0, \tau];
(\tau, 2 \tau]; \ldots $ the velocity fields are assumed to be statistically
independent and have the same statistics. This implies that the velocity
field loses memory at the prescribed instants $ t = n \tau ,$
where $ n = 0, \pm 1, \pm 2, \ldots .$
This velocity field cannot be considered as a
stationary  velocity field for small times $ \sim \tau ,$ however, it
behaves like a stationary field for $ t \gg \tau .$
Note that the fields $ b_{j}(s, {\bf x}') $ and
$ Q_{ij}(t, s, {\bf x}, {\bf x}') $ are statistically
independent because the field $ b_{j}(s, {\bf x}') $ is determined in
the time interval $ (- \infty, s] ,$ whereas the function
$ Q_{ij}(t, s, {\bf x}, {\bf x}') $ is defined on the
interval $ (s, t] .$ Due to a renewal, the velocity
field as well as its functionals $ b_{j}(s, {\bf x}') $ and
$ Q_{ij}(t, s, {\bf x}, {\bf x}') $ in these two time
intervals are statistically independent.
Now we make a change of variables $ ({\bf x},{\bf x}') \to ({\bf x},{\bf x}'
= {\bf z}+{\bf x}) $ in Eq. (\ref{T8}), i.e.,
$ \tilde Q_{ij}(t,s,{\bf x},{\bf x}') = \tilde Q_{ij}(t,s,{\bf x},{\bf z}
+{\bf x}) = Q_{ij}(t,s,{\bf x},{\bf z}) .$
The Fourier transformation in Eq. (\ref{T8}) yields
\begin{eqnarray*}
b_{i}(t, {\bf x}) = & & \int \int Q_{ij}(t,s,{\bf x},{\bf k})
\exp(i {\bf k} \cdot {\bf z}) \,d {\bf k}
\\
& \times & \int b_{j}(s, {\bf q}) \exp[i {\bf q} \cdot ({\bf z}+{\bf x})]
\,d {\bf q} \,d {\bf z} \; .
\end{eqnarray*}
Since $ \delta({\bf k} + {\bf q}) =  (2 \pi)^{-3} \int
\exp[i ({\bf k} + {\bf q}) \cdot {\bf z})] \,d {\bf z} ,$
we obtain that
\begin{eqnarray}
b_{i}(t, {\bf x}) &=& (2 \pi)^{3} \int Q_{ij}(t,s,{\bf x},-{\bf
q}) b_{j}(s, {\bf q})
\nonumber\\
& & \times \exp(i {\bf q} \cdot {\bf x}) \,d{\bf q} \; .
\label{T13}
\end{eqnarray}
In Eq. (\ref{T13}) the function $ Q_{ij}(t,s,{\bf x},-{\bf q}) $ is
given by
\begin{eqnarray}
Q_{ij}(t,s,{\bf x},-{\bf q}) &=& (2 \pi)^{-3} \int Q_{ij}(t,s,{\bf
x},{\bf z})
\nonumber\\
& & \times \exp(i {\bf q} \cdot {\bf z}) \,d{\bf z} \; .
\label{C2}
\end{eqnarray}
Substituting $ \tilde Q_{ij}(t,s,{\bf x},{\bf x}')
= Q_{ij}(t,s,{\bf x},{\bf z}) $ in Eq. (\ref{T8}) and taking into account
that $ {\bf x}' = {\bf z} + {\bf x} $ we obtain
\begin{eqnarray}
b_{i}(t,{\bf x}) = \int Q_{ij}(t,s,{\bf x},{\bf z})  b_{j}(s,{\bf z} + {\bf x})
\,d {\bf z} \; .
\label{C3}
\end{eqnarray}
Equation (\ref{C2}) can be rewritten in the form
\begin{eqnarray}
(2 \pi)^{3} Q_{ij}(t,s,{\bf x},&-&{\bf q}) \exp(i {\bf q} \cdot
{\bf x}) = \int Q_{ij}(t,s,{\bf x},{\bf z})
\nonumber\\
& & \times \exp[i {\bf q} \cdot ({\bf z} + {\bf x})] \,d{\bf z}
\; . \label{C4}
\end{eqnarray}
The right hand sides of Eqs. (\ref{C3}) and (\ref{C4}) coincide when
$ {\bf b}(s,{\bf z} + {\bf x}) = {\bf e} \, \exp[i {\bf q} \cdot ({\bf z}
+ {\bf x})] ,$ where $ {\bf e} $ is a unit vector.
Thus, a particular solution (\ref{C3}) of Eq. (\ref{T1}) with the initial
condition $ {\bf b}(s, {\bf x}') = {\bf e} \, \exp(i {\bf q} \cdot {\bf x}') $
coincides in form with the integral (\ref{C4}). On the other hand,
a solution of Eq. (\ref{T1}) is given by Eq. (\ref{T2}).
Substituting the initial condition $ {\bf b}(s,\bec{\zeta}) = {\bf e}
\, \exp(i {\bf q} \cdot \bec{\zeta}) = {\bf e} \, \exp[i {\bf q} \cdot ({\bf x}
+ (2 D_{m})^{1/2} {\bf w})] $ into Eq. (\ref{T2}) we obtain
\begin{eqnarray}
b_{i}(t,{\bf x}) &=& M_{\bec{\zeta}} \{J(t,s,\bec{\zeta}) \tilde
G_{ij}(t,s,\bec{\zeta}) e_{j} \,
\nonumber\\
& & \times \exp[i {\bf q} \cdot ({\bf x} + (2 D_{m})^{1/2} {\bf
w})] \} \; . \label{C5}
\end{eqnarray}
Comparing Eqs. (\ref{C3})-(\ref{C5}) we get
\begin{eqnarray}
Q_{ij}(t,s,{\bf x},-{\bf q}) &=& (2 \pi)^{-3} M_{\bec{\zeta}}
\{J(t,s,\bec{\zeta}) \tilde G_{ij}(t,s,\bec{\zeta}) \,
\nonumber\\
& & \times \exp[i (2 D_{m})^{1/2} {\bf q} \cdot {\bf w}] \} \; .
\label{C6}
\end{eqnarray}
Now we rewrite Eq. (\ref{C6}) using Feynman-Kac formula (\ref{T5}).
The result is given by
\begin{eqnarray}
Q_{ij}(t,s,{\bf x},-{\bf q}) &=& (2 \pi)^{-3} M_{\bec{\xi}}
\{G_{ij}(t,s,\bec{\xi}(t,s)) \,
\nonumber\\
& & \times \exp[i {\bf q} \cdot \bec{\xi}^{\ast}] \} \;,
\label{C7}
\end{eqnarray}
where $ \bec{\xi}^{\ast} = \bec{\xi} - {\bf x} .$
Substituting Eq. (\ref{C7}) into Eq. (\ref{T13}) we obtain
\begin{eqnarray}
b_{i}(t, {\bf x}) &=& \int M_{\bec{\xi}} \{G_{ij}(t,s,\bec{\xi})
\, \exp[i {\bf q} \cdot \bec{\xi}^{\ast}] b_{j}(s, {\bf q}) \}
\nonumber\\
& & \times \exp(i {\bf q} \cdot {\bf x}) \,d{\bf q} \; .
\label{C8C}
\end{eqnarray}
The Fourier transformation in Eq. (\ref{C8C}) yields Eq. (\ref{A5}).
The above derivation proves that the assumption
(\ref{CC8}) is correct for a Wiener path $ \bec{\xi} .$
In order to derive equation for the second-order correlation function
$ \Phi_{ij}(t, {\bf x}, {\bf y}) =  \langle b_{i}(t, {\bf x})
b_{j}(t, {\bf y}) \rangle $ we use Eq. (\ref{C8C}), where
the angular brackets $ \langle \cdot \rangle $
denote the ensemble average over the random velocity field.
After the Fourier transformation we obtain
\begin{eqnarray}
\Phi_{ij}(t, {\bf x}, {\bf y}) = (2 \pi)^{-6} \int \int
P_{ijpl}(\tau, {\bf x}, {\bf y}, {\bf k}_{1}, {\bf k}_{2})
\nonumber\\
\times \exp[i ({\bf k}_{1} \cdot {\bf x} + {\bf k}_{2} \cdot {\bf
y})] \biggl[\int \int \Phi_{pl}(s, {\bf x}', {\bf y}')
\nonumber \\
\times \exp[- i ({\bf k}_{1} \cdot {\bf x}' + {\bf k}_{2} \cdot
{\bf y}')] \,d {\bf x}' \,d {\bf y}' \biggr] \,d {\bf k}_{1} \,d
{\bf k}_{2} \;, \label{A48}
\end{eqnarray}
where
\begin{eqnarray}
P_{ijpl}(&\tau&, {\bf x}, {\bf y}, {\bf k}_{1}, {\bf k}_{2}) =
M_{\bec{\xi}} \{ \langle G_{ip}({\bf x}) G_{jl}({\bf y})
\nonumber \\
& & \times \exp[i ({\bf k}_{1} \cdot \bec{\xi}^{\ast}({\bf x}) +
{\bf k}_{2} \cdot \bec{\xi}^{\ast}({\bf y}))] \rangle \} \;,
\label{AA48}
\end{eqnarray}
$ G_{ij}({\bf x}) \equiv G_{ij}(\tau, \bec{\xi}({\bf x})) $ and
$ \tau = t - s .$
For a homogeneous and isotropic random flow Eq. (\ref{A48}) reads
\begin{eqnarray}
\Phi_{ij}(t, {\bf r}) &=& \int \int P_{ijpl}(\tau, - {\bf q}, {\bf
r}) \exp[i {\bf q} \cdot ({\bf r} - {\bf r}')]
\nonumber \\
& & \times  \Phi_{pl}(s, {\bf r}') \,d {\bf r}' \,d {\bf q} \;,
\label{A49}
\end{eqnarray}
where $ {\bf r} = {\bf y} - {\bf x} ,$
\begin{eqnarray}
P_{ijpl}(\tau, - {\bf q}, {\bf r}) &=& M_{\bec{\xi}} \{ \langle
G_{ip}({\bf x}) G_{jl}({\bf y})
\nonumber \\
& & \times  \exp(i {\bf q} \cdot \bec{\tilde \xi}) \rangle \}  \;
\label{A50}
\end{eqnarray}
and $ \bec{\tilde \xi} = \bec{\xi}^{\ast}({\bf y}) -
\bec{\xi}^{\ast}({\bf x}) .$ The Fourier transformation of Eq. (\ref{A49})
yields Eq. (\ref{A51}).

\section{Derivation of Eq. (\ref{D8})}

Now we use the model of the random velocity field with a small
correlation time. We expand the functions $ \bec{\xi}^{\ast} $
and $ G_{ij}(\tau,\bec{\xi}) $ in Taylor series of small
time $ \tau .$ Then an expression for the
function $ P_{ijpl}(\tau,{\bf r}, i \bec{\nabla}) $ reads:
\begin{eqnarray}
P_{ijpl}(\tau,{\bf r}, i \bec{\nabla}) &=& \delta_{ip} \delta_{jl}
+ \tau B_{ijpl} + \tau U_{ijplm} \nabla_{m}
\nonumber \\
& & +  \tau D_{ijplmn} \nabla_{m} \nabla_{n} + \ldots \;,
\label{D1}
\end{eqnarray}
where
\begin{eqnarray}
D_{ijplmn} &=& (1 / 2 \tau) M_{\bec{\xi}} \{ \langle
\tilde \xi_{m} \tilde \xi_{n} G_{ip}({\bf x}) G_{jl}({\bf y}) \rangle \} \;,
\label{D2} \\
U_{ijplm}(r) &=& \tau^{-1} [\delta_{jl} M_{\bec{\xi}} \{\langle g_{ip}({\bf x})
\xi_{m}^{\ast}({\bf y}) \rangle \}
\nonumber \\
& & + \delta_{ip} M_{\bec{\xi}} \{\langle g_{jl}({\bf x})
\xi_{m}^{\ast}({\bf y}) \rangle \}
\nonumber \\
&-& (1/2) M_{\bec{\xi}}
\{\langle g_{ip}({\bf x}) g_{jl}({\bf y}) \tilde \xi_{m} \rangle \}]  \;,
\label{D3} \\
B_{ijpl}(r) &=& \tau^{-1} M_{\bec{\xi}} \{\langle g_{ip}({\bf x})
g_{jl}({\bf y}) \rangle \} \;,
\label{D4}
\end{eqnarray}
and $ \quad G_{ij} = \delta_{ij} + g_{ij} $ and $ M_{\bec{\xi}}
\{\langle g_{ij} \rangle \} = 0 .$
Thus an equation for the second-order correlation function for a
magnetic field in a random velocity field with a small yet
finite  correlation time reads:
\begin{eqnarray}
{\partial \Phi_{ij} \over \partial t} &=&  [B_{ijpl} + U_{ijplm}
\nabla_{m}
\nonumber \\
& & + D_{ijplmn} \nabla_{m} \nabla_{n}] \Phi_{pl}(t,{\bf r}) \; .
\label{D5}
\end{eqnarray}
Now we consider a random velocity field with a Gaussian
statistics. This assumption allows us to calculate the tensors
$ D_{ijplmn} ,$ $ \quad U_{ijplm} $ and $ B_{ijpl} .$  We omit the lengthy
algebra and present the final results:
\begin{eqnarray}
D_{ijplqn} &=& D_{ijplqn}^{(1)} + D_{ijplqn}^{(2)} + D_{ijplqn}^{(3)}
\nonumber \\
& & + 2 D_{m} \delta_{qn} \delta_{ip} \delta_{jl} \;,
\label{B1} \\
D_{ijplmn}^{(1)} &=& 2 \tau \{ \tilde f_{mn} + {\rm St}^{2}
[(\nabla_{s} f_{kn}) (\nabla_{k} f_{ms})
\nonumber \\
& &- \tilde f_{sk} (\nabla_{s} \nabla_{k} f_{mn})] \} \delta_{ip}
\delta_{jl} \;,
\label{B2} \\
D_{ijplmn}^{(2)} &=& (1/2) \tau {\rm St}^{2}
[(\nabla_{k} f_{im}) (\nabla_{p} f_{nk})
\nonumber \\
& & - (\nabla_{k} f_{mn}) (\nabla_{p} f_{ik})
\nonumber \\
& &+ 2 \tilde f_{ms} (\nabla_{s} \nabla_{p} f_{in})] \delta_{jl} \;,
\label{B3} \\
D_{ijplmn}^{(3)} &=& (1/2) \tau {\rm St}^{2}
[(\nabla_{p} f_{im}) (\nabla_{l} f_{jn})
\nonumber \\
& & - \tilde f_{mn} (\nabla_{p} \nabla_{l} f_{ij})] \;,
\label{B4} \\
B_{ijpl} &=& - 2 \tau \{ (\nabla_{p} \nabla_{l} f_{ij}) - {\rm St}^{2}
[(\nabla_{k} \nabla_{s} f_{ij}) (\nabla_{p} \nabla_{l} f_{ks})
\nonumber \\
& & + 2 (\nabla_{p} \nabla_{m} f_{is}) (\nabla_{l} \nabla_{s} f_{jm})] \} \;,
\label{B5} \\
U_{ijplm} &=& 4 \tau \{ (\nabla_{p} f_{im}) \delta_{jl} + {\rm St}^{2}
[((\nabla_{k} f_{is}) (\nabla_{p} \nabla_{s} f_{km})
\nonumber \\
& & + (\nabla_{p} f_{sk}) (\nabla_{k} \nabla_{s} f_{im})
\nonumber \\
& & - (\nabla_{s} f_{km}) (\nabla_{k} \nabla_{p} f_{is}))
\delta_{jl}
\nonumber \\
& & + 2 ((\nabla_{k} f_{jm}) (\nabla_{p} \nabla_{l} f_{ik})
\nonumber \\
& & + (\nabla_{l} f_{km}) (\nabla_{k} \nabla_{p} f_{ij}))] \} \;,
\label{B6}
\end{eqnarray}
where $ {\rm St} = \tau u_{d} / l_{d} $ is the Strouhal number,
$ \tilde f_{mn} = f_{mn}(0) - f_{mn}({\bf r}) ,$
and we changed $ \tau \to 2 \tau $ in order to compare the obtained
results with those for the $\delta$-correlated in time approximation
for a random velocity field. Here the small terms of the order of
$ \sim O({\rm St}^{4}) $ are being neglected. In Eqs. (\ref{B1})-(\ref{B6})
we took into account a commutative symmetry in every pair of the
following indexes: $ (i,j); $ $ (p,l) $ and $ (m,n) .$
The latter is due to a symmetry of the following tensors: $ r_{ij} ,$
$ \Phi_{pl} $ and $ \nabla_{m} \nabla_{n} .$
In Eqs. (\ref{B1})-(\ref{B6}) we assumed also that the form of the tensor
$ \tilde f_{mn} $ is given by $ \tilde f_{mn} = C_{mnps} r_{p} r_{s} ,$ where
$ C_{mnps} $ is an arbitrary constant tensor. This satisfies for the model
of the velocity field (\ref{D6}) with $ F(r) = 1 - r^{2} .$

Now we seek a solution for the second moment of the magnetic field
in the form of Eq. (\ref{D7}).
Multiplying Eq. (\ref{D5}) by $ r_{ij} $ and using Eq. (\ref{D7})
we obtain the equation for the correlation function $ W(t,r) =
\langle b_{\bf r}(t,{\bf x}) b_{\bf r}(t,{\bf y}) \rangle .$
This equation is given by Eq. (\ref{D8}). For the derivation
of Eq. (\ref{D8}) we used the following identities
\begin{eqnarray}
\hat D W & \equiv & r_{ij} D_{ijplmn} \nabla_{m} \nabla_{n}
\Phi_{pl} = {2 \tau \over 3} [r^{2} W'' + 8 r W'
\nonumber \\
& & + {\sigma_{_{\xi}}  \over 4} (2 r^{3} W''' + 31 r^{2} W'' + 12
r W')
\nonumber \\
& & + {3 \over \Pr} (W'' + 4 W / r)] \;,
\label{B7} \\
\hat B W & \equiv & r_{ij} B_{ijpl} \Phi_{pl} = {4 \tau \over 3}
[2 r W' + 5 W
\nonumber \\
& & + {\sigma_{_{\xi}}  \over 4} (r W' + 5 W)] \;,
\label{B8} \\
\hat U W & \equiv & r_{ij} U_{ijplm} \nabla_{m} \Phi_{pl} = {2
\tau \over 3} \{ - 6 r W'
\nonumber \\
& & + {\sigma_{_{\xi}}  \over 4} [r^{2} W'' + (29 / 2) r W'] \} \;
. \label{B9}
\end{eqnarray}
Equations (\ref{B7})-(\ref{B9}) are derived
by means of Eqs. (\ref{D6}),  (\ref{B1})-(\ref{B6})
and we also used the following identities:
\begin{eqnarray}
\nabla_{n} \Phi_{pl} = {1\over 2} \biggl[\biggl(W'' - {W' \over r}
\biggr) P_{pl} r_{m} + {W' \over r} (4 \delta_{pl} r_{m}
\nonumber \\
- \delta_{pm} r_{l} - \delta_{lm} r_{p})\biggr] \;,
\label{B10} \\
\nabla_{m} \nabla_{n} \Phi_{pl} = {1\over 2} \biggl[(r W''')
P_{pl} r_{mn} + \biggl(W'' - {W' \over r} \biggr)
\nonumber\\
\times (P_{mn} P_{pl} + 4 P_{pl} r_{mn} - P_{pm} r_{ln} - P_{lm}
r_{pn}
\nonumber \\
- P_{pn} r_{lm} - P_{ln} r_{pm} + 2 r_{plmn}) + {W' \over r} (4
\delta_{pl} \delta_{mn}
\nonumber \\
- \delta_{pm} \delta_{ln} - \delta_{lm} \delta_{pn}) \biggr] \; .
\label{B11}
\end{eqnarray}
The corresponding derivatives for $ f_{pl} $ coincide with Eqs.
(\ref{B10}) and (\ref{B11}) after the change $ W(r) \to (1 / 3)
F(r) .$ Note that for $ F(r) = 1 - r^{2} $ the following
identities are valid: $ F'' - F' / r = 0 $ and $ F''' = 0 .$
Turbulent magnetic diffusion is determined by function $ \hat D W
= r_{ij} D_{ijplmn} \nabla_{m} \nabla_{n} \Phi_{pl}(t,{\bf r}) .$
The latter depends on the field of Lagrangian trajectories $
\bec{\xi} $ [see Eqs. (\ref{D2}) and (\ref{D5})]. Due to a finite
correlation time of a random velocity field $ \langle
(\bec{\nabla} \cdot \bec{\xi})^{2} \rangle \not=0 $ even if the
velocity field is incompressible. Indeed, $ \langle (\bec{\nabla}
\cdot \bec{\xi})^{2} \rangle \approx (4/9) {\rm St}^{4} =
\sigma_{_{\xi}}^{2} .$ Thus the parameter $ \sigma_{_{\xi}} $
describes the compressibility of the field of Lagrangian
trajectories. The latter results in a change of the dynamics of
magnetic fluctuations. Thus, the equation for the correlation
function $ W(t,r) $ is given by Eq. (\ref{D8}).

\end{document}